\documentclass[10pt,letterpaper]{article}
\pdfoutput=1
\usepackage{cogsci}
\usepackage{placeins}
\usepackage{natbib}
\usepackage{graphicx}
\usepackage{tikz}
\usetikzlibrary{positioning,shadows,arrows}
\usepackage{amsmath, amssymb}
\usepackage{algorithm}
\usepackage[noend]{algpseudocode}

\newcommand\numberthis{\addtocounter{equation}{1}\tag{\theequation}}

\title{Simple Trees in Complex Forests: Growing Take The Best by Approximate Bayesian Computation}
 \author{\textbf{Eric Schulz}$^1$ (e.schulz@cs.ucl.ac.uk), \textbf{Maarten Speekenbrink}$^1$ (m.speekenbrink@ucl.ac.uk) \&\\ \textbf{Bj\"orn Meder}$^2$ (meder@mpib-berlin.mpg.de)\\
$^1$Department of Experimental Psychology, University College London, Gower Street, London WC1E 6BT, UK\\
$^2$Max Planck Institute for Human Development, Lentzeallee 94, 14195 Berlin, Germany}

\begin{document}

\maketitle

\begin{abstract}
How can heuristic strategies emerge from smaller building blocks? We propose Approximate Bayesian Computation (ABC) as a computational solution to this problem. As a first proof of concept, we demonstrate how a heuristic decision strategy such as Take The Best (TTB) can be learned from smaller, probabilistically updated building blocks. Based on a self-reinforcing sampling scheme, different building blocks are combined and, over time, tree-like non-compensatory heuristics emerge. This new algorithm, coined \emph{Approximately Bayesian Computed Take The Best} (ABC-TTB), is able to recover data that was generated by TTB, leads to sensible inferences about cue importance and cue directions, can outperform traditional TTB, and allows to trade-off performance and computational effort explicitly.\medskip
\end{abstract}

\section{Introduction}
How can heuristic strategies emerge from smaller building blocks? Consider the heuristic \emph{Take The Best} (TTB; Gigerenzer \& Goldstein, 1996) \nocite{gigerenzer1996reasoning}as an example: TTB is a lexicographic, non-compensatory strategy to decide which of two objects scores higher on an unobserved criterion variable, based on accessible pieces of information (cues). The heuristic looks up cues sequentially (lexicographically), with the order of cues determined by their importance (predictive value). TTB is non-compensatory because once a cue discriminates between the two options, a decision is made irrespectively of the value of the other cues. Although relatively simple, TTB has been shown to outperform more sophisticated models such as regression models,  nearest neighbour classifiers, and classification and regression trees, across many data sets \citep{gigerenzer2009homo}. 

A key question is how such a successful but domain-specific strategy is learned \citep{gigerenzer2011heuristic}. According to the \emph{adaptive toolbox} metaphor, there are three building blocks that can be used to implement decision heuristics. \emph{Search rules} dictate how information is acquired, for example looking up cues ordered by their predictive value. \emph{Stopping rules} determine when to stop searching, for example as soon as a cue discriminates between the two options. Lastly, \emph{decision rules} determine the actual decision after search has stopped, for example deciding which of the two objects scores higher on the target criterion, depending on the cue direction (i.e. whether a cue is positively or negatively related to the criterion). Considering TTB, we see that all of the building blocks are present. 

A viable theory of heuristic decision making should also explain how heuristics strategies like TTB are learned, that is, how they emerge during learning from their constituent building blocks. One idea is that a decision maker starts off with some prior intuitions about which cues might be helpful, and then learns  from experience which cue orders and directions are successful. We formalize this idea using \emph{Approximately Bayesian Computed Take The Best} (ABC-TTB), a computational approach towards learning TTB adaptively by approximate Bayesian computation. When heuristic building blocks are fed into it, successful decision strategies are learned on the fly, paralleling a Bayesian reinforcement learning algorithm. Note that our approach critically differs from approaches that model \emph{strategy selection} through reinforcement learning mechanisms \citep[e.g.,][]{rieskamp2006ssl} as we do not use strategies  (e.g., TTB vs. weighting-and-adding) but building blocks from within a model class (e.g., different TTB variants) as the unit of reinforcement. Our approach is based on the acceptance and rejection of simple simulations that explore the usefulness of proposal models, emerging from probabilistically updated building blocks. \\

We show that ABC-TTB:
\begin{enumerate}
\item recovers TTB when TTB provides the task structure.\vspace{-5.5px}
\item generates sound inference based on learned cue order and directions.\vspace{-5.5px}
\item can outperform traditional TTB and other models.\vspace{-5.5px}
\item can balance performance and computational effort.
\end{enumerate}
 This is a first proof of concept of how Take The Best can emerge from smaller building blocks.

\section{Approximate Bayesian Computation}
Bayesian inference concerns updating prior beliefs in light of observed data. Given a prior distribution $\pi(\theta)$ reflecting our initial beliefs about an unknown parameter $\theta$, the data $\mathcal{D}$ affect the posterior belief $p(\theta|\mathcal{D})$ only via the likelihood $p(\mathcal{D}|\theta)$:
\begin{align*}
p(\theta|\mathcal{D})&=\frac{p(\mathcal{D}|\theta)\pi(\theta)}{\int_{\theta} p(\mathcal{D}|\theta)\pi(\theta) \text{d} \theta}\\
& \propto p(\mathcal{D}|\theta)\pi(\theta)\numberthis \label{eqn}
\end{align*}

Bayesian inference thus models subjective beliefs as a mixture of prior assumptions and incoming data, providing a coherent framework to model inference under uncertainty. Accordingly, the approach has been used to model cognition in many domains \citep{oaksford2007bayesian}. Since full Bayesian inference is frequently intractable \citep[e.g., due to the curse of dimensionality, see][]{bellman1961adaptive}, powerful sampling mechanisms have been developed that approximate the posterior distribution by means of Markov chain Monte Carlo methods \citep{gilks2005markov}. 
Here, we focus on an alternative approach to approximate posterior inference: \emph{Approximate Bayesian Computation} \citep[ABC;][]{turner2012tutorial}. ABC is an approximative method that, instead of computing the required likelihood directly, substitutes it with surrogate simulations of a model \citep{csillery2010approximate}, and then checks whether the output of these simulations comes close enough to the actual data. This simple mechanism of model simulations and reality checks provides a useful tool to approximate posteriors and has been applied to many scenarios in which the true underlying likelihood is hard to assess. 
 
A formal description of ABC is shown in Algorithm~\ref{alg:abc}. The algorithm samples a proposed parameter from a prior $\pi(\theta)$ and plugs the proposal into a model $\mathcal{M}$ in order to simulate an output $\mathbf{y^*}$ from the given data $\mathcal{D}$. It then calculates a summary statistic $\mathcal{S}$ of both the simulated data and the real data and accepts proposals that have produced a summary that is close enough (measured by a distance $\rho$) to the summary of the real data, allowing for some error $\epsilon$. The proximity of the two summary statistics is normally estimated by the difference of the statistics $\delta$, for example the Euclidean distance between two means. A more intuitive explanation of this algorithm is that an agent has some subjective beliefs about how the world works and repeatedly checks whether or not these beliefs can, on average, produce similar patterns to the ones observed. If they can, the model proposals are accepted and (possibly) reinforced. If they cannot, the proposals are rejected. Even though this algorithm is based on a computationally simple reinforcement scheme, it has been shown to provide reasonable approximate posteriors in many scenarios, in particular when inferring tree-like structures (Beaumont, Zhang, \& Balding, 2002)\nocite{beaumont2002approximate}.
\begin{algorithm}
\caption{Approximate Bayesian Computation}
\label{alg:abc}
\begin{algorithmic}
\Require $\pi(\theta)$; model $\mathcal{M}$; data $\mathcal{D}=(\mathbf{X}, \mathbf{y})$; tolerance $\epsilon$, required sample size $\eta$
\While{$j<\eta$}\\
 \hspace{5mm}\textbf{Sample} $\theta^{*}\sim \pi(\theta)$\\
\hspace{5mm}\textbf{Simulate} $\mathbf{y}^{*}=\mathcal{M}_{\theta^{*}}(\mathbf{X})$\\
\hspace{5mm}\textbf{Calculate} $\delta=\rho \big(\mathcal{S}(\mathbf{y}),\mathcal{S}(\mathbf{y}^{*})\big)$
\If {$ \delta\leq \epsilon$}
    \State set $\theta_j = \theta^*$ and $j = j + 1$
\Else
            \State reject $\theta^*$
\EndIf
\EndWhile
\textbf{end while}
\end{algorithmic}
\end{algorithm}
\section{ABC-TTB: Growing heuristic strategies}
We next show how ABC can be applied to the problem of learning heuristics from smaller building blocks. More specifically, our ABC-TTB algorithm can yield a similar tree-like structure as TTB, but learns the structure of the tree (cue order and cue directions) on the fly while it makes decisions and receives feedback regarding the usefulness of the model proposals.\footnote{Even though TTB is not commonly labelled as a tree, the model can be represented as a tree structure, as shown in Figure~\ref{fig:ttbtruth}.} Thus, ABC-TTB randomly generates proposal trees and checks how well these trees perform in a subset of the data. Successful cues are reinforced through a kind of Bayesian reinforcement learning algorithm \citep{poupart2010bayesian}. This way, ABC-TTB starts out with small building blocks, combines and tests them, and --over time-- heuristic structures emerge that reflect the learning experience. 
 
\begin{figure}[htb!]
  \centering
    \includegraphics[scale=0.35]{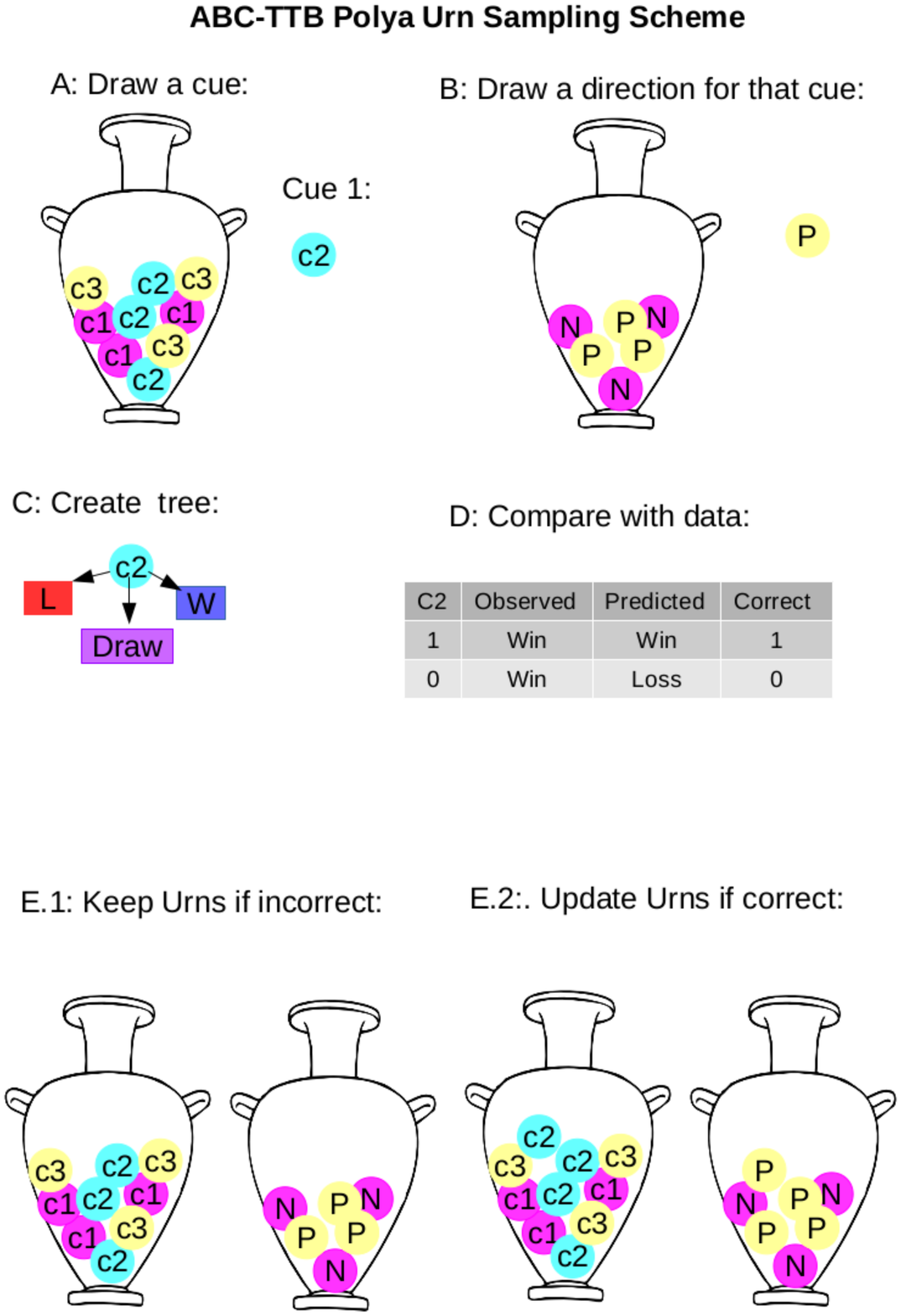}
    \caption{Sketch of ABC-TTB's sampling scheme. c1-c3 are three example cues; N and P are negative or positive weights.}
\label{fig:scheme}
\end{figure}

Figure~\ref{fig:scheme} uses a Polya urn sampling scheme to illustrate the reinforcement-based learning mechanism underlying ABC-TTB. The model learns through updating distributions over cue importance and cue directions, from which model proposals are generated and tested. Figure~\ref{fig:scheme}a shows the urn representing three available cues, $c1$ to $c3$. A cue is sampled from the current distribution; in this case $c2$ is sampled from a uniform distribution over the cues. Next a cue direction is sampled from an independent urn associated with this particular cue (Figure~\ref{fig:scheme}b). From these two building blocks a proposal model is generated (Figure~\ref{fig:scheme}c) and tested against the data (Figure~\ref{fig:scheme}d). If the cue correctly predicts which object scores higher on the target criterion, the sampled cue and direction are reinforced by putting a $c2$ ball into the cue urn and a $P$ ball into the urn associated with $c2$'s direction. If the prediction is wrong, the proposal is rejected and no reinforcement takes place. This process is repeated several times, and over time the distributions resulting from reinforcing successful cue combinations make it more likely that successful model proposals are generated.  

Now consider a probabilistic implementation of this learning process. Associated with each cue is a Beta prior ($B(1,1)$) that induces an implicit likelihood for each cue to generate a successful prediction. From these priors, a probability is independently sampled for each cue, which are used to generate a proposal tree by normalizing these probabilities (so they sum up to 1) and then sampling a cue according to these probabilities (similar to Figure~\ref{fig:scheme}a). The drawn cue is the top node of the tree (i.e. the most important cue). The second cue is drawn from the remaining cues by first renormalizing the associated probabilities and then sampling accordingly. Thus, cues are drawn in order without replacement from the set of all cues to construct a lexicographic decision tree. Consider a situation with three cues, where $c1$ has a probability of 0.8 to make a correct prediction, $c2$ a probability of 0.6, and $c3$ a probability of 0.2 (all sampled from the corresponding Beta priors). To draw the top node of the tree, these probabilities are normalized to sum to 1. For example, the probability that $c1$ is sampled as the top node is $.8/(.8+.6+.2) = .5$, while the second and third cue have probabilities of $.375$ and $.125$ respectively. Imagine that $c1$ was drawn as the top node. The remaining probabilities for $c2$ and $c3$ are then re-normalized to $.75$ and $.25$, respectively, and the second cue is sampled. This process continues until all cues have been sampled. Associated to each cue is a second Beta prior which models the probability of the cue direction (i.e., whether it is positively or negatively related to the outcome). The resulting beta-binomial distribution model assigns to each cue the probability to have a positive direction. The cue direction is independently sampled for each cue and attached to the nodes to generate the proposal tree. 

The proposal tree is then used to make predictions for a randomly sampled subset of the available data (e.g., the training sample in cross-validation). The size of the sampled subset can be varied by a parameter $0<\phi\leq1$, which means that the subset has to be more than 0\% and can contain up to 100\% of the data. If more than (1-$\epsilon)\times100$\% of the predictions are correct, the proposal is accepted and the cues involved within that tree, as well as their directions, get updated by adding a success to their posterior beta-binomial-distribution in proportion to how often they have actually made a difference (how often they successfully discriminated between the cues in the sub-sample). As the Beta distributions are updated based on a cue's success, more successful cues are sampled more often as the top node in  proposal trees than less successful cues. If less than (1-$\epsilon)\times100$\% of the predictions are correct, the proposal tree is rejected and no update happens for the cues or their directions. This whole process repeats until $\eta$ successful proposal trees have been generated. At this point, the updated Beta distributions are approximations to the posterior distributions of cues and directions. 

ABC-TTB makes predictions for a new paired comparison based on the posterior distributions of cues and directions resulting from the learning process. The posteriors are  used to generate $\omega$ proposal trees and the final decision is given by the modal prediction from all sampled proposal trees' predictions. In summary, the ABC-TTB algorithm constitutes a computational solution to how cue importance and cue directions are learned simultaneously through a simple reinforcement process. The algorithm shows how a heuristic can emerge from combining and reinforcing its building blocks, resulting in an  algorithm that is able to find and utilize simple trees within the complex forest of all possible trees.  

\section{Recovering TTB}
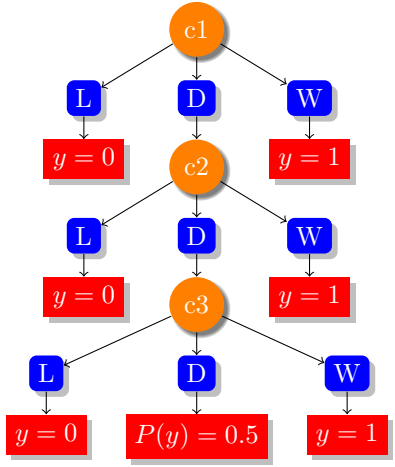
\begin{figure}
\centering
\begin{tikzpicture}[
    fact/.style={rectangle, draw=none, rounded corners=1mm, fill=blue, drop shadow,
        text centered, anchor=north, text=white},
    state/.style={circle, draw=none, fill=orange, circular drop shadow,
        text centered, anchor=north, text=white},
    leaf/.style={rectangle, draw=none, fill=red, drop shadow,
        text centered, anchor=north, text=white},
    level distance=0.3cm, growth parent anchor=south
]
\node (State00) [state] {c1} [->]
child{
        node (Fact01) [fact] {L}
child{
                            node (State03) [leaf] {$y=0$}
                        }
        }
    child{
        node (Fact01) [fact] {D}
        child{
            node (State01) [state] {c2}
           child{
        node (Fact04) [fact] {L}
child{
                            node (State05) [leaf] {$y=0$}
                        }
        } 
           child{
                node (Fact02) [fact] {D}
                child{ [sibling distance=2cm]
                    node (State02) [state] {c3}
child{
        node (Fact31) [fact] {L}
child{
                            node (State33) [leaf] {$y=0$}
                        }
        }
                    child{[sibling distance=9cm]
                        node (Fact03) [fact] {D}
                        child{
                            node (leafguess) [leaf] {$P(y) = 0.5$}
                        }
                    }
child{
        node (Fact99) [fact] {W}
child{
                            node (State99) [leaf] {$y=1$}
                        }
        }
}}
child{
        node (Fact100) [fact] {W}
child{
                            node (State100) [leaf] {$y=1$}
                        }
        }
}}
child{
        node (Fact01) [fact] {W}
child{
                            node (State03) [leaf] {$y=1$}
                        }
        }

;
        
\end{tikzpicture}
\caption{Underlying TTB-model generating the data set. c4 is not shown as it is not predictive.}
\label{fig:ttbtruth}
\end{figure}

A first sanity check is to see whether ABC-TTB can recover the true underlying model from a data set that has been generated by the TTB heuristic. As TTB is a subset of all possible models (cue orders and cue directions) captured by ABC-TTB, it should be able to recover it from simulated data. We simulated a paired-comparison task with four cues, according to the TTB model shown in Figure~\ref{fig:ttbtruth}. For each cue, the two objects in a paired comparison could either be identical (a draw, with probability .5), or the cue could be present in one object and absent in the other (a win, or loss, each with probability .25). In this environment the outcome of a decision is determined by the first discriminating (winning or losing) cue, just like in the original TTB. The cues were ordered by importance as $c1$, $c2$, $c3$, and $c4$. If the third cue $c3$ did not discriminate (a draw), the outcome was determined randomly. As a result, the fourth cue, $c4$, was uninformative in the simulation. The decision strategy generating this data set is identical to TTB, i.e. it can be seen as a comparison between two different items A and B, where a win means that item A has something that item B has not and an outcome of $y=1$ means that item A is better than item B. 

\begin{figure}[htb!]
  \centering
    \includegraphics[scale=0.4]{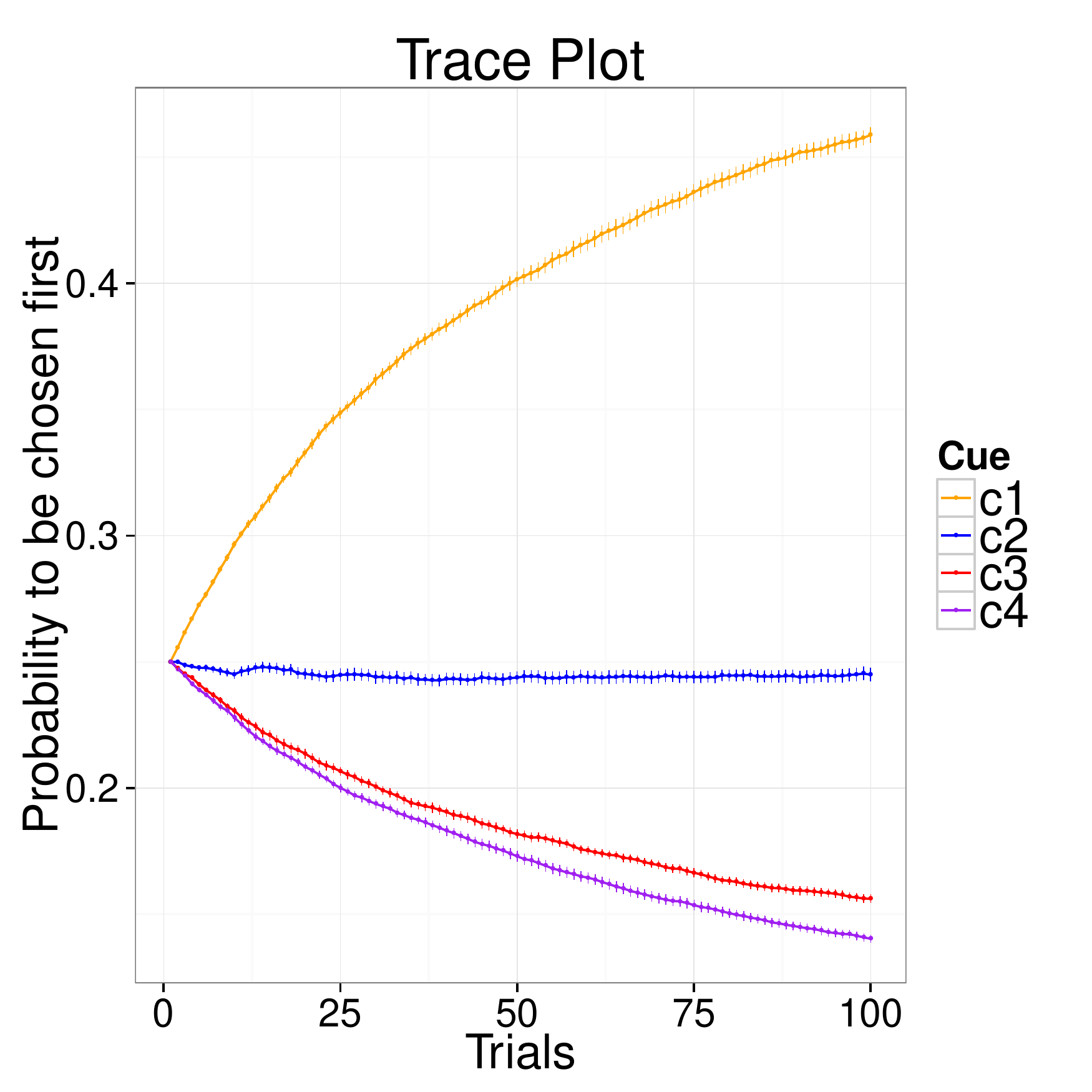}
    \caption{Trace plot of probability to be chosen first over 100 accepted proposals, averaged over 100 simulations. Legend is ordered by final cue importance. Bars represent standard errors.}
\label{ttbrecover}
\end{figure}

A data set of size $n=1000$ was generated and the ABC-TTB algorithm was run with $\epsilon=0.1$ and $\phi=0.1$ until $\eta=100$ proposed trees were accepted. Figure~\ref{ttbrecover} shows the cue importance over time. The traces were calculated by dividing each Beta's posterior mean by the total sum of posterior means of all cues. This then equals the probability of a cue to be chosen as the top node in the tree and can act as a proxy variable for a cue's overall importance. Notice that the actual magnitude of that probability is not as important as the recovered order. The simulation was repeated 100 times and the probabilities were averaged per step. It can be seen that --even after a few accepted proposals-- the mode tree (the most likely tree) is the same as the one generating the data. Shortly after the start of the simulation, the cue order grows to the correct order and less important cues are chosen less frequently, exactly as expected given the underlying structure. Additionally, the uninformative cue $c4$ approached a probability of 0. 

\section{Inference and performance for real data}
\begin{figure}[!htb]
    \centering
    \includegraphics[scale=0.45]{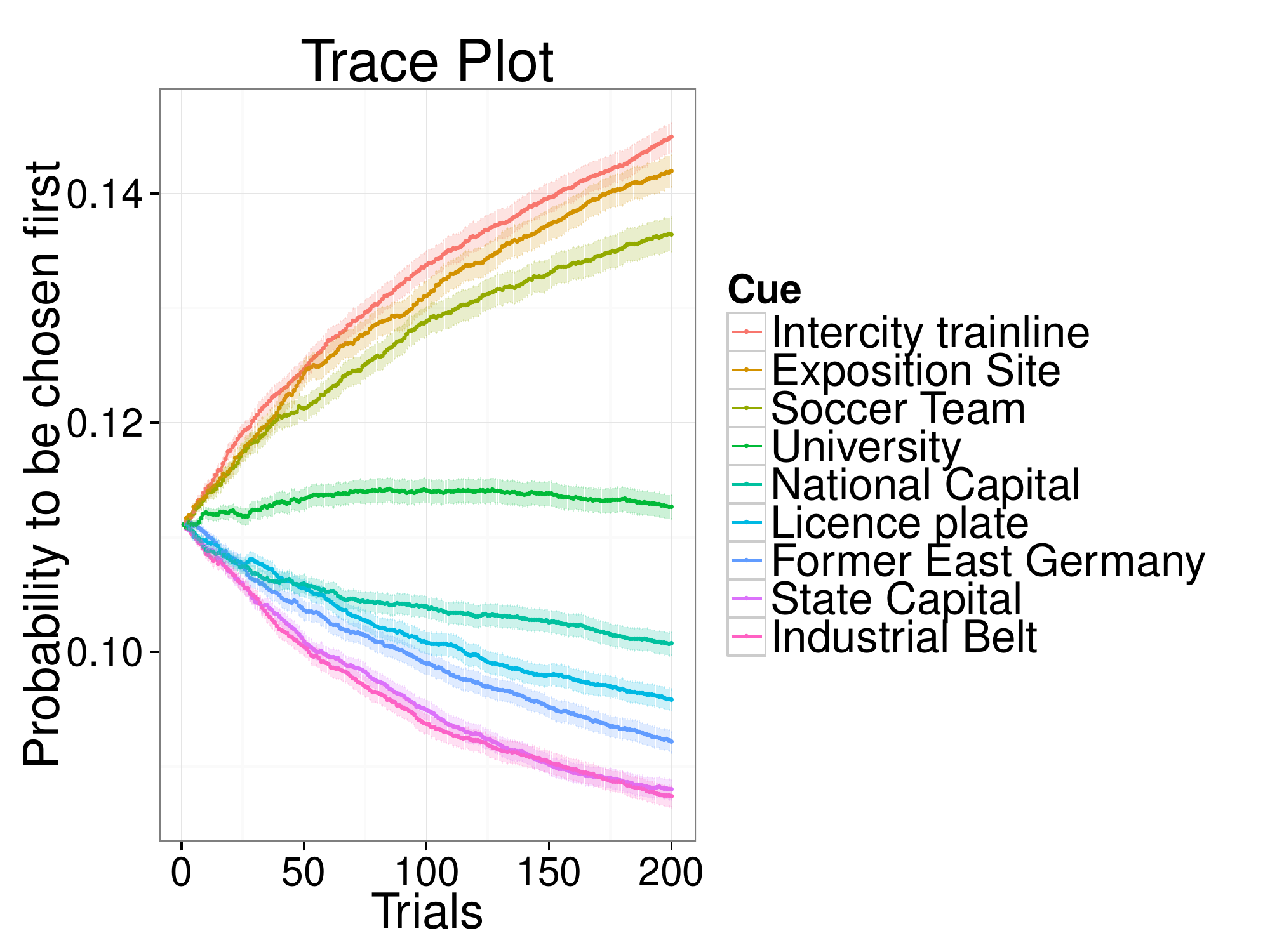}
\caption{Trace plot of probability to be chosen first over 200 accepted proposals, averaged over 100 simulations. Legend is ordered by final cue importance. }
\label{fig:citytrace}
\end{figure}

Next, we checked whether or not ABC-TTB yields sensible and interpretable inferences in a real-world data set. For this, we applied our algorithm to the classic city size data, which has been frequently used in previous research \citep[e.g.,][]{gigerenzer2009homo} and contains the population size of different German cities and whether or not they have an intercity train line, an exposition site, a soccer team, a university, are the national capital, have their own license plate, are located in the former east Germany, are a state capital, and have their own industrial belt . This data set contains 81 objects (German cities) with 9 different binary variables that can be used to decide which of two cities has more citizens (the unobserved target criterion). Figure~\ref{fig:citytrace} shows the trace plots for the city size simulation ($\epsilon=0.35$, $\phi=0.01$)\footnote{We used different parameters as the data set was bigger and contained more variables. See next section on a more detailed treatment of hyper-parameters.} over 200 accepted proposal trees. Again, the results of ABC-TTB look sensible, finding that having a intercity train line,  an exposition site, a major league soccer team, and a university are the four most important cues. This cue order correspond more to a frequency-adjusted validity taking into account how often a given cue can be utilized \citep[cf.][]{newell2004search}. 

\begin{figure}[t!]
   \centering
    \includegraphics[scale=0.4]{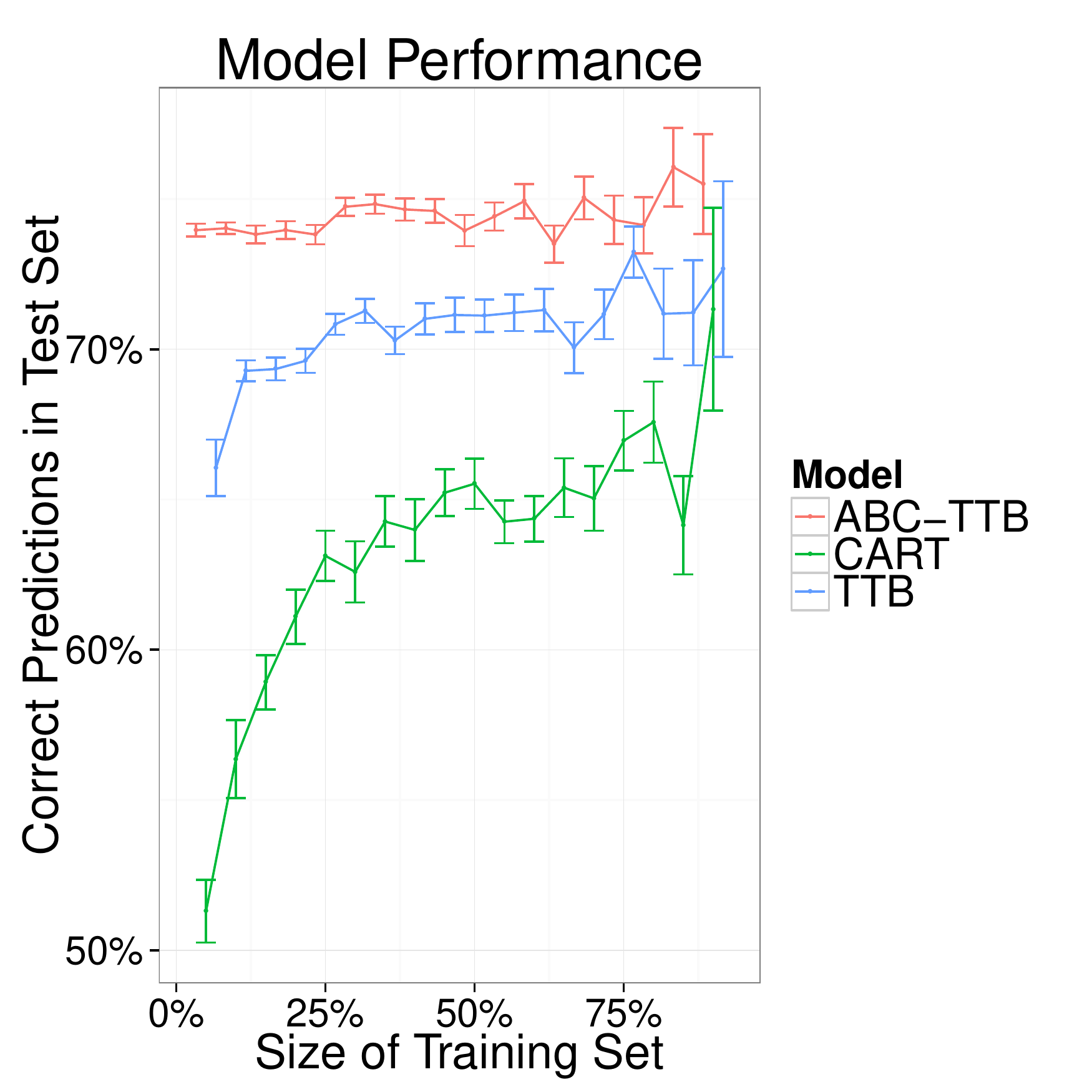}
\caption{Performance of TTB, ABC-TTB, and CART in the city size data set. Performance increases with training set. ABC-TTB performs best overall.}
\label{fig:perfcity} 
\end{figure}

Next, we compared the performance of ABC-TTB to the performance of classic TTB within the city size data set. Additionally, we also tested classification trees (CART), another tree-based classification algorithm. For the comparison, we split up the data into learning sets of $s=[5,10,\dots,90]$\%, fitted TTB (using the classic cue validity), CART, and ABC-TTB ($\epsilon=0.5$, $\rho=0.1$) to this data and then assessed their predictive accuracy in the remaining test set. This procedure was repeated 50 times for every learning set size and the results were averaged over all trials. Results can be seen in Figure~\ref{fig:perfcity}.
ABC-TTB consistently outperformed classic TTB, which itself performed better than CART \citep[c.f.][]{gigerenzer2009homo}. This is due to the fact that ABC-TTB learns the tree structure by sub-sampling items from the learning set and makes predictions based on aggregations over possible outcomes. As it can be seen as an ensemble of different trees, ABC-TTB will naturally generate less prediction errors than classic TTB. The posterior of ABC-TTB defines a distribution over possible trees, some of which will generate divergent predictions. Therefore, predictive variances can be calculated, which is important for future research on exploration-exploitation problems (Schulz, Konstantinidis \& Speekenbrink, 2015)\nocite{schulz2015learning}.

\section{Performance-effort trade-off}
Another benefit of ABC-TTB is that performance and computational effort can be traded-off against each other explicitly. Assessing this trade-off lies at the heart of approaches that frame a learning agent as exhibiting a meta-reasoning mechanisms sensitive to the costs of cognition (Gershman, Horvitz \& Tenenbaum, 2015)\nocite{gershman2015computational}, a hypothesis that can only be evaluated by making the accuracy-effort-off explicit.

\begin{figure*}[htb!]
 \centering
  \includegraphics[scale=0.52]{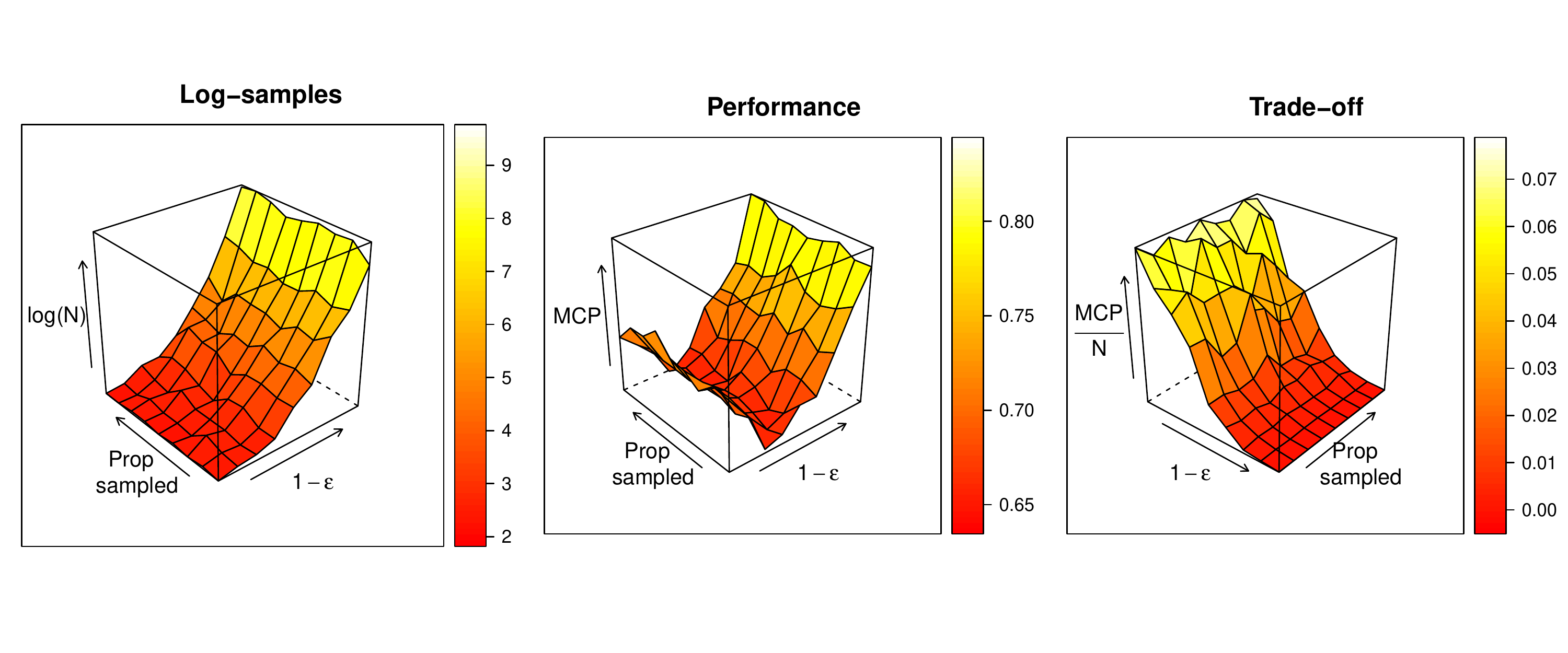}
 \vspace{-1em}
\caption{Assessed  computational effort (log-number of proposal trees, N), performance (mean correct predictions, MCP), and trade-off (proportion correct predictions per proposal generated) for different $\epsilon$-$\phi$-combinations. Number of samples go up exponentially when bigger proportions are sampled and only high quality proposals are accepted. Performance also goes up, but there is a diminishing returns effect whereby increasingly less additional correct predictions are generated per generated proposal.}
\label{fig:perftradeoff}
\end{figure*}

As $(1-\epsilon)$ defines the proportion of predictions that have to be correct in order for a proposal tree to be accepted, decreasing $\epsilon$ will lead to more rejections of proposed trees and to acceptance of only higher quality proposals. The same holds true for the proportion of sampled data points $\phi$. The higher the proportion of the sampled data set, the better the proposed tree has to be, and the longer it takes to find good trees. We can adjust these parameters explicitly to gauge the interaction between performance and computational effort. Therefore, we created all 81 possible combinations of $\epsilon=\{0.1,0.2,\dots,0.9\}$ and $\rho=\{0.1,0.2,\dots,0.9\}$ and applied the resulting models to the scenario in which TTB had generated the data as in the ``Recovering TTB'' analysis above. We tracked the number of proposed samples as a proxy variable of computational effort, averaged over 100 trials. Additionally, we tracked how well ABC-TTB described the generated data overall, measured by the mean proportion of correct predictions it would generate within that sample. Lastly, we calculated the ratio of the mean correct predictions (MCP) per proposal generated. Results are shown in Figure~\ref{fig:perftradeoff}.

The leftmost plot in Figure~\ref{fig:perftradeoff} shows that the computational effort (measured by the number of generated proposal trees, plotted on a logarithmic scale) seems to increase more than exponentially (exponentially on a log-scale) when the proportion of samples and the quality of the to-be-accepted proposals increases. For the performance of the resulting models (shown in the middle), we can see that tuning the $\epsilon$ and $\phi$-parameters can increase performance by up to 20\%. Most importantly, there seems to be a diminishing returns property as shown in the rightmost plot, in the sense that one would have to spend more and more proposal samples for an ever smaller increase in prediction accuracy. Taking these results together, we see that there seems to be a clear trade-off between effort and performance as number of samples increases super-exponentially, but performance increases not as steeply. Using ABC-TTB, we can change the parameters directly and test this accuracy-effort trade-off explicitly thereby gaining deeper insights into when good predictions might only need few samples \citep{vul2014one}.
\FloatBarrier

\section{Conclusion}
We have introduced Approximate Bayesian Computation as a method to let heuristic strategies emerge from smaller building blocks. ABC assesses the posterior of a model by sanity checking simulated proposals. We therefore think it could be a plausible approach to model how diverse strategies emerge over time. As a first proof of concept, we have shown how simple TTB-like trees can emerge from smaller building blocks (distributions over cues and their weights). Our new model, coined ABC-TTB, can recover TTB when data was generated by it, produces sensible inference and predictions in a real world data set, and allows to trade off computational effort and performance explicitly, consistently generating simple trees within the complex forest of all possible trees.

This is a first step towards unpacking the heuristic toolbox and future work will focus on extending our approach to the problem of letting more diverse building blocks or even different strategies emerge. Another venue for future research concerns the psychological plausibility of the proposed learning mechanism. We think that dynamic scenarios such as active \citep{parpart2015active} or causal learning \citep{morais2014causal} are a plausible way to test sequential predictions of heuristic models, especially given that ABC-TTB generates decisions from the beginning of the learning process and provides measures of uncertainty for every prediction. Given the large number of cognitive models (Schulz, Speekenbrink \& Shanks, 2014)\nocite{schulz2014predict}, we believe that probing how strategies emerge from compositional building blocks, and how simple structures are derived from approximative computation, is a promising endeavour.

\section{Acknowledgements}

ES is supported by the UK Centre for Doctoral Training in Financial Computing. He would like to thank the MPIB-ABC research group for hosting him multiple times. BM was supported by grant ME 3717/2-2. Code available at: github.com/ericschulz/TTBABC
{\footnotesize
\bibliographystyle{apa-good}
\bibliography{Bibo}}
\end{document}